\documentclass[british, a4paper]{article}

\usepackage[british]{babel}
\usepackage{hyperref}
\usepackage{fullpage}
\usepackage{babel}
\usepackage{mathtools}
\usepackage{amsmath}
\usepackage{amsthm}
\usepackage{amssymb}
\usepackage{authblk}
\usepackage[capitalise]{cleveref}
\usepackage{pgfplots}
\usepackage{subcaption}
\usepackage{eurosym}
\usepackage{nth}
\usepackage{graphicx}

\pgfplotsset{width=10cm,compat=1.9} 

\newtheorem{thm}{Theorem}

\providecommand{\keywords}[1]{\textbf{\textit{Index terms---}} #1}

\begin{document}

\title{Fast Calibration of Two-Factor Models for Energy Option Pricing}

\author[1]{Emanuele Fabbiani}
\author[1]{Andrea Marziali}
\author[1]{Giuseppe De Nicolao}

\affil[1]{Department of Electrical, Computer and Biomedical Engineering, University of Pavia}

\maketitle

\begin{abstract}
Energy companies need efficient procedures to perform market calibration of stochastic models for commodities. If the Black framework is chosen for option pricing, the bottleneck of the market calibration is the computation of the variance of the asset. Energy commodities are commonly represented by multi-factor linear models, whose variance obeys a matrix Lyapunov differential equation. In this paper, analytical and numerical methods to derive the variance are discussed: the Lyapunov approach is shown to be more straightforward than \textit{ad-hoc} derivations found in the literature and can be readily extended to higher-dimensional models. A case study is presented, where the variance of a two-factor mean-reverting model is embedded into the Black formulae and the model parameters are calibrated against listed options. The analytical and numerical method are compared, showing that the former makes the calibration 14 times faster. A Python implementation of the proposed methods is available as open-source software on GitHub.

\vspace{15px}
\end{abstract}

\section{Introduction}\label{sec:intro}
Since the 1990s, the liberalization of energy markets fostered the adoption of derivatives: nowadays, accurate pricing models are of interest for every energy company. The Black formula \cite{Black1973, Black1976} provides the no-arbitrage price of a European vanilla option on a future, while pricing exotic derivatives often requires Monte Carlo simulations \cite{glasserman2013monte}. In both cases, it is assumed that the underlying assets are described by stochastic processes: a first issue is then choosing the most appropriate model. It is known that the Geometric Brownian Motion adopted in the Black formula does not capture fundamental features of energy commodities. A number of diffusion processes specific to energy assets were then proposed \cite{lucia2002electricity, schwartz1997stochastic, weron2007modeling, pilipovic2007energy} and, for their flexibility, multi-factor models became a popular choice \cite{Schwartz2000, fanelli2019seasonality}. 

A second but equally important problem is calibrating the model, that is tuning its parameters so that it reflects the behaviour of an asset. Two types of calibration exist. Historical calibration aims at fitting the model to the spot prices of the asset: several methods, including Kalman filter \cite{MartinBarlow2004} and Monte Carlo Markov Chain \cite{guerini2019mcmc}, were proposed. Historically-calibrated models, however, may be biased and fail in achieving arbitrage-free prices. As models are often used to value derivatives traded over the counter, a procedure which guarantees arbitrage-free pricing is required. Market calibration fits the stochastic model to liquid derivatives, usually European options, thus providing no-arbitrage guarantees \cite{weber2011american, cui2017full, nastasi2020smile}.  

This study is motivated by a practical issue: the need, by energy companies, for efficient market calibration of multi-factor models. As common practice in the industry, the Black framework is adopted, but the underlying is described by a two-factor model. Stochastic volatility (SV) is not considered, as the calibration of SV models remains demanding despite recent developments \cite{cui2017full, floc2018adaptive}. 

The bottleneck of the market calibration is the computation of the variance of the underlying. For many stochastic processes, including some two- and three-factor models, analytical formulae for the variance are available, but their derivation is often complex and convoluted \cite{Kiesel2009, Schwartz2000}. The main contribution of this paper is to show that such results can be obtained by a systematic procedure, amenable to symbolic computation. We exploit the fact that, when the underlying is represented by a linear stochastic system, its variance obeys a Lyapunov matrix differential equation \cite{gajic2008lyapunov}, irrespective of the model order. Using the Lyapunov method, we derive the analytical expression of the variance of the Log-spot price mean reverting to generalised Wiener process two-factor model (LMR-GW) \cite{MartinBarlow2004}. As a further example, we present an alternative and simpler derivation for the two-factor model by Schwartz at al. \cite{Schwartz2000} 

Both the numerical and analytical solutions of the Lyapunov equation are discussed: in either case, the key point is the computation of a Gramian integral, which can be performed analytically or through the numerical calculation of a matrix exponential. A comparison between the two solutions is carried out in terms of computational speed. This is a crucial factor, as market calibration is implemented by optimization programs, which repeatedly evaluate the pricing formula. We finally showcase the proposed calibration procedure on European options collected from the EEX electricity market and the TTF gas market. 

The paper is organized as follows: \cref{sec:black-framework} recalls the fundamentals of the Black framework, and \cref{sec:mean-reverting-models} concisely presents the considered models. The variance derivation through the Lyapunov equation is provided in \cref{sec:variance-derivation}, with a discussion on computational efficiency. \cref{sec:market-calibration} describes the calibration procedure and its results on the test cases. Finally, in  \cref{sec:conclusion} some concluding remarks end the paper. In \cref{appendix-schwartz}, the Schwartz model is used as an example to demonstrate the simplicity of the Lyapunov approach.

\section{Pricing framework}\label{sec:black-framework}
In energy markets, underlyings of options are usually averages of futures on a given period (month, quarter or year). The Black formula is a widely accepted framework to price vanilla options on futures \cite{Black1973, Black1976}.
Assume that a future $s$ behaves like a Geometric Brownian Motion (GBM) with zero mean and standard deviation $\sigma$,
\begin{equation}
ds\left(t\right)=\sigma s\left(t\right)dw\left(t\right),
\end{equation}
where $w\left(t\right)$ is a Wiener process. Then, the no-arbitrage prices $c$ of a European call option and $p$ of a European put option on $s$ are:
\begin{subequations}
\label{eq:bl-price}
\begin{align}
c & =e^{-rT}\left(S_{0}N\left(d_{1}\right)-KN\left(d_{2}\right)\right)\\
p & =e^{-rT}\left(KN\left(-d_{2}\right)-S_{0}N\left(-d_{1}\right)\right) \\
d_{1} & =\frac{\ln\left(S_{0}/K\right)+\left(\sigma^{2}/2\right)T}{\sigma\sqrt{T}}\\
d_{2} & =d_{1}-\sigma\sqrt{T},
\end{align}
\end{subequations}
$S_{0}$ being the price of $s$ at time $t=0$, when the option is traded, $K$ the strike, $T$ the maturity and $N$ the cumulative probability distribution of a standard Gaussian variable. The listed price $S_{0}$ takes into account the seasonal behaviour embedded in the corresponding maturity.

Recalling that $\sigma\sqrt{t}$ is the standard deviation of the GBM process, this term can be interpreted as the uncertainty on the log-return of the underlying at maturity. One could then devise alternatives models for the underlying, and then plug their standard deviation into the Black formulae. A rigorous discussion of this approach is provided in several previous works \cite{Schwartz2000, Kiesel2009, clewlow2000energy}. 

Suppose that the variance of log-return of the underlying at time $t$ is given by a positive function of time $p$. Then, $d_{1}$ and $d_{2}$ can be written as

\begin{equation}
d_{1} = \frac{\ln\left(S_{0}/K\right)+\frac{1}{2}p\left(T\right)}{\sqrt{p\left(T\right)}}, \quad
d_{2} = d_{1}-\sqrt{p\left(T\right)}.
\label{eq:bl-with-model-variance}
\end{equation}

The function $p$ is evaluated at a single time instant, so that pricing does not take into account the evolution of the underlying future after the maturity. This simplification is sensible in most energy markets, because the option maturity coincides with (or is very close to) the beginning of the delivery period of the future. The coincidence between the maturity of the option and the underlying future also justifies the adoption of models usually designed for spot prices for pricing options written on futures \cite{clewlow2000energy}. 
The Black formula can be slightly simplified by noting that, nowadays, proxies for the risk-free interest rate are close to zero, or even negative. In the following it is assumed $r=0$. 

\section{Mean-reverting models}\label{sec:mean-reverting-models}

GBM does not take into account mean-reversion. Prices of energy commodities and related futures follow a long-term trend: if, for whatever reason, they get away from it, they tend to be pushed back within a short time span \cite{pindyck1999long}. 

The simplest mean-reverting model is the Ornstein-Uhlenbeck process. Still, the volatility of an Ornstein-Uhlenbeck process is asymptotically constant, thus failing to capture the ever-increasing uncertainty over long maturities. To overcame this issue, two-factor models were proposed \cite{Schwartz2000, MartinBarlow2004}: of particular interest is the Log-spot price mean reverting to generalised Wiener process model (LMR-GW) \cite{MartinBarlow2004}. In LMR-GW, the log-return $x_1$ of the future $s$ follows an Ornstein-Uhlenbeck process, accounting for the short-term variations, while the long-term drift $x_2$ follows a GBM:
\begin{subequations}
\label{eq:two-factor-model}
\begin{align}
x_1(t) &= \ln s\left(t\right)\\
dx_{1}\left(t\right) &= \lambda\left(x_{2}\left(t\right)-x_{1}\left(t\right)\right)dt+\sigma_{1}dw_{1}\left(t\right)\\
dx_{2}\left(t\right) &= \mu dt+\sigma_{2}dw_{2}\left(t\right),
\end{align}{}
\end{subequations}
where $\lambda, \sigma_1, \sigma_2$ and $\mu$ are scalar parameters, $w_{1}$ and $w_{2}$ are independent Wiener processes.

To exploit the LMR-GW model for pricing, an expression for its variance is required. The problem is not new in the literature: in particular, it is worth mentioning the analytical solution worked out by Schwartz et al. \cite{Schwartz2000} for a differently formulated second-order model. However, their derivation was targeted to a specific model, so that its extension to other two-factor or higher-order processes is not straightforward. 
In the next section, leveraging the theory of linear stochastic systems, we show that the variance of a wide class of models, including \eqref{eq:two-factor-model}, can be computed in a systematic way.

\section{Variance derivation by Lyapunov equation} \label{sec:variance-derivation}
Consider a continuous-time scalar stochastic process $y$ described by the state-space representation
\begin{subequations}
\label{eq:linear-system}
\begin{align}
dx\left(t\right)&=Ax\left(t\right)dt+Bdw\left(t\right)\\
y\left(t\right)&=Cx\left(t\right), \label{eq:linear-system-output}
\end{align}
\end{subequations}
where $A$, $B$ and $C$ are matrices of suitable dimensions, $x$ is the $n$-dimensional state, $w$ an $m$-dimensional Wiener process, such that $\mathrm{E}\left[w(t) w(\tau)^T\right] = S(|t-\tau|), S = S^T > 0$, and y a $q$-dimensional output. To complete the description of the system, initial values for both the expected value and the variance of the state are required: $\bar{x}_{0} \coloneqq \mathrm{E}\left[x\left(0\right)\right]$, $P_{0} \coloneqq  \mathrm{Var}\left[x\left(0\right)\right]$, $P_{0} = P_{0}^T > 0$.
Let $P\left(t\right) = \mathrm{Var}\left[x\left(t\right)\right]$ denote the covariance matrix of the system state. From \eqref{eq:linear-system}, it follows that $P$ satisfies the Lyapunov matrix differential equation:
\begin{equation}
\label{eq:lyapunov-general}
\frac{dP\left(t\right)}{dt}=AP\left(t\right)+P\left(t\right)A^{T}+BSB^{T},
\end{equation}
under the initial condition $P\left(0\right)=P_{0}$.
A solution to this equation is given by the matrix version of the Lagrange formula \cite{kwakernaak1972linear, gajic2008lyapunov}:
\begin{equation}
\label{eq:lyapunov-solution}
P\left(t\right)=e^{At}P_{0}e^{A^{T}t}+\int_{0}^{t}e^{A(t-z)}BSB^{T}e^{A^{T}\left(t-z\right)}dz, 
\end{equation}
where $e^{M} = \text{exp}(M)$ denotes the matrix exponential of $M$. Moreover, the variance of the output is:
\begin{equation}
\mathrm{Var}[y(t)] = \mathrm{Var}[Cx(t)] = CP(t)C^T.
\label{eq:output-variance}
\end{equation}

\subsection{Variance of LMR-GW two-factor model}
In the LMR-GW process \eqref{eq:two-factor-model}, the drift parameter $\mu$ does not affect the variance. The system can be rearranged in the form of \eqref{eq:linear-system} by letting $x(t) = \begin{bmatrix}x_1(t) &  x_2(t)\end{bmatrix}^T$, $dw(t) = \begin{bmatrix}dw_1(t) &  dw_2(t)\end{bmatrix}^T$ and
\begin{equation}
A=\begin{bmatrix}-\lambda & \lambda\\
0 & 0
\end{bmatrix}, \quad
B=\begin{bmatrix}\sigma_{1} & 0\\
0 & \sigma_{2}
\end{bmatrix}, \quad
S=\begin{bmatrix}1 & 0\\
0 & 1
\end{bmatrix}.
\end{equation}
We do not need \eqref{eq:linear-system-output} as the output equals the first state. Let $P$ and $P_{0}$ denote the state covariance and its initial value, respectively:
\begin{equation}
P\left(t\right)=\begin{bmatrix}P_{11}\left(t\right) & P_{12}\left(t\right)\\
P_{21}\left(t\right) & P_{22}\left(t\right)
\end{bmatrix}
\quad
P_{0}=\begin{bmatrix}p_{11} & p_{12}\\ p_{21} & p_{22}
\end{bmatrix}.
\end{equation}
Observe that the exponential of $At$ con be easily computed:
\begin{equation}
	e^{At}=\begin{bmatrix}e^{-\lambda t} & 1-e^{-\lambda t}\\ 0 & 1	\end{bmatrix}.
\end{equation}
By applying \eqref{eq:lyapunov-solution}, after some algebraic manipulation, the analytical solution of the Lyapunov equation \eqref{eq:lyapunov-general} is found:
\begin{subequations}
\label{eq:analytical-solution}
\begin{align}
P_{11}\left(t\right) & =\left(p_{11}-2p_{12}+p_{22}-\frac{\sigma_{1}^{2}+\sigma_{2}^{2}}{2\lambda}\right)e^{-2\lambda t}+ \nonumber \\
 & +2\left(p_{12}-p_{22}+\frac{\sigma_{2}^{2}}{\lambda}\right)e^{-\lambda t}+\sigma_{2}^{2}t+\frac{\sigma_{1}^{2}-3\sigma_{2}^{2}}{2\lambda}+p_{22} \\
P_{12}\left(t\right)=P_{21}\left(t\right) & =\left(p_{12}-p_{22}+\frac{\sigma_{2}^{2}}{\lambda}\right)e^{-\lambda t}+\sigma_{2}^{2}t+p_{22}-\frac{\sigma_{2}^{2}}{\lambda} \\
P_{22}\left(t\right) & =\sigma_{2}^{2}t+p_{22}.
\end{align}
\end{subequations}
As the log-price of the underlying is represented by $x_1$, for pricing purposes the only relevant term is $P_{11}$. 

Despite the variance of the LMR-GW model was already known in the literature, the application of the Lyapunov equation makes the derivation far easier. Another example can be found in \cref{appendix-schwartz}, where we derive the variance of the Schwartz model \cite{Schwartz2000}. 

The method based on the Lyapunov equation is general and can be applied to every linear stochastic system which follows the assumptions stated for \eqref{eq:linear-system}. General procedures to compute the matrix exponential are discussed in the classical book by Hall \cite{Hall2015}, but it must be noted that our problem calls for the solution of the matrix exponential to be written as a function of time. Even for models with a relatively low dimension, finding an analytical expression for $e^{At}$ can be challenging or even impossible. Therefore, alternative methods are of interest.

\subsection{Numerical solution of the Lyapunov equation}
The matrix exponential $e^{At}$ is key to solve the Lyapunov equation: if an analytical solution is not available, one could choose a numerical approximation. An efficient procedure relies on the following theorem \cite{Chen1995}.

\begin{thm}[Exponential of triangular matrix]
\label{thm:matrix-exponential}
Let $M_{11}$, $M_{12}$ and $M_{22}$ be matrices of suitable dimensions.
Let
\begin{equation}
\begin{bmatrix}F_{11} & F_{12}\\
\boldsymbol{0} & F_{22}
\end{bmatrix}=\exp\left(\begin{bmatrix}M_{11} & M_{12}\\
\boldsymbol{0} & M_{22}
\end{bmatrix}h\right),
\end{equation}
where $\boldsymbol{0}$ is a null matrix of appropriate dimensions. Then, the following identities hold:
\begin{equation}
F_{11} = e^{M_{11}h}, \quad 
F_{22} = e^{M_{22}h}, \quad
F_{12} = \int_{0}^{h}e^{M_{11}\left(h-z\right)}M_{12}e^{M_{22}z}dz.
\end{equation}
\end{thm}

In \cref{thm:matrix-exponential} and in the following, the dependency of $F_{11}$, $F_{12}$, and $F_{22}$ on time is omitted for readability. We want to apply \cref{thm:matrix-exponential} to solve the Lagrange formula \eqref{eq:lyapunov-solution} and find the state covariance matrix $P$. To this goal, we define $M_{11}$, $M_{12}$, and $M_{22}$ so that
\begin{equation}
\begin{bmatrix}F_{11} & F_{12}\\
\boldsymbol{0} & F_{22}
\end{bmatrix}=\exp\left(\begin{bmatrix}A & BSB^{T}\\
\boldsymbol{0} & -A^{T}
\end{bmatrix}t\right)
\end{equation}
to get:
\begin{equation}
F_{11} = e^{At}, \quad
F_{22} = e^{-A^{T}t}, \quad
F_{12} = \int_{0}^{t}e^{A\left(t-z\right)}BSB^{T}e^{-A^{T}z}dz.
\end{equation}
With a few more manipulations,
\begin{equation}
\int_{0}^{t}e^{A\left(t-z\right)}BSB^{T}e^{A^{T}\left(t-z\right)}dz=F_{12}F_{22}^{-1}.
\end{equation}
The solution of the differential Lyapunov matrix equation can
thus be expressed as
\begin{equation}
P\left(t\right)=F_{11}P_{0}F_{22}^{-1}+F_{12}F_{22}^{-1}=\left(F_{11}P_{0}+F_{12}\right)F_{22}^{-1}.
\label{eq:numerical-solution}
\end{equation}
Note that $F_{22}$ is always invertible because it is the result of a matrix exponential.

The above procedure is general-purpose: it can be applied on every linear stochastic system which follows the assumptions stated for \eqref{eq:linear-system}.

\subsection{Numerical and analytical solution: comparison} \label{subsec:num-and-analytical-comparison}
Generality and ease of implementation make the numerical solution \eqref{eq:numerical-solution} attractive. However, in practical applications, computational efficiency is also an important factor. The main use cases for the variance are pricing and market calibration of stochastic models, which both require it to be evaluated repeatedly. In the daily routine of energy companies, calibration is run over multiple markets, amplifying the importance of small differences in execution time. This motivates us to perform comparative tests.

Both the numerical and the analytical approach were implemented using Python and its SciPy package \cite{scipy2019}. Matrix exponential was calculated by the \texttt{expm} function of the \texttt{scipy.linalg} module, which employs the Pad\'e approximant, improved with scaling and squaring methods \cite{Al-Mohy2009}. The implementation of both the numerical and the analytical solution, as well as other related tools, is available in an open-source Python package \cite{fabbiani2019vanilla, fabbiani2020vanilla}. The hardware was a commercial off-the-shelf personal computer, running an Intel i5 3340M two-core CPU and 16 GB of RAM. 

An LMR-GW model with market-calibrated parameters was considered. A time window of 30 days was set, within which the variance was computed at $M$ time instants, with $M$ ranging from 1 to 30,000. Each run was repeated 10 times and the median of the CPU times was taken. Speedups were derived as the ratio of the CPU time of the numerical solution to that of the analytical one. 

The data presented in \cref{tab:CPU-time-and-speedups} and \cref{fig:CPU-time-against-numer-of-evaluations} suggests that a speedup of 40 can be achieved if an analytical formula for the variance of the underlying is available.

\begin{table}
\centering
\begin{tabular}{r@{\extracolsep{0pt}.}lr@{\extracolsep{0pt}.}lr@{\extracolsep{0pt}.}lr@{\extracolsep{0pt}.}l}
\hline 
\multicolumn{2}{c}{\textbf{Evaluations}} & \multicolumn{2}{c}{\textbf{Analytical}} & 
\multicolumn{2}{c}{\textbf{Numerical}} & \multicolumn{2}{c}{\textbf{Speedup}}\tabularnewline
\hline 
\multicolumn{2}{c}{1,000} & 0&0259 & 0&9956 & 39&41\tabularnewline
\multicolumn{2}{c}{3,000} & 0&0738 & 3&0751 & 41&77\tabularnewline
\multicolumn{2}{c}{5,000} & 0&1239 & 5&0849 & 41&45\tabularnewline
\multicolumn{2}{c}{10,000} & 0&2885 & 10&5574 & 37&44\tabularnewline
\multicolumn{2}{c}{20,000} & 0&4958 & 20&8032 & 42&17\tabularnewline
\multicolumn{2}{c}{30,000} & 0&7394 & 31&4745 & 42&83\tabularnewline
\hline
\end{tabular}\caption{CPU time and speedup. All the figures are in seconds.}
\label{tab:CPU-time-and-speedups}
\end{table}

\begin{figure}
    \centering
	\begin{tikzpicture}
	\begin{semilogyaxis}[
		xlabel=Evaluations,
		ylabel=Time (s),
		legend pos=south east,
		height=5cm,
        grid=major,
        grid style={thin, dotted, black!20},
	]
	\addplot[orange, mark=square, thick] table [y=Analytical,x=Evaluations]{data/speedup.dat};
	\addlegendentry{Analytical}
	\addplot[blue, mark=o, mark options=solid, dashed] table [y=Numerical,x=Evaluations]{data/speedup.dat};
	\addlegendentry{Numerical}
	\end{semilogyaxis}
	\end{tikzpicture}
\caption{CPU time against number of evaluations - semilog scale. \label{fig:CPU-time-against-numer-of-evaluations}}
\end{figure}

\section{Application to market calibration}\label{sec:market-calibration}
To be effective in pricing exotic derivatives or options traded over the counter, stochastic models must have their parameters tuned. As mentioned in \cref{sec:intro}, we focus on market calibration, which is preferred for pricing tasks. In order to appreciate the difference with historical calibration, we provide a comparison in \cref{sec:experimental-results}. For sake of clarity, in the following we consider only LMR-GW, but the procedure can be readily applied to every linear model.

It is common practice to tune the stochastic models on liquid instruments, such as European options: in this view, market calibration is an inverse problem of pricing. Let us freeze the market at a specific instant and let $O_{i}$ be the prices of $n$ European options, $i=1,...,n$, written on the same underlying; let also $\hat{O}_{i}$ be the price of the $i$-th option given by the Black formula \eqref{eq:bl-price}, embedding the variance of an LMR-GW model  - \eqref{eq:bl-with-model-variance} and \eqref{eq:analytical-solution}:
\begin{align}
\hat{O}_i &= \begin{cases}
    e^{-rT_i}\left(S_{0, i}N\left(d_{1, i}\right)-K_iN\left(d_{2, i}\right)\right) & \text{if $i$ is a call } \\
    e^{-rT_i}\left(K_iN\left(-d_{2, i}\right)-S_{0, i}N\left(-d_{1, i}\right)\right) & \text{if $i$ is a put }
  \end{cases} \\
d_{1, i} &= \frac{\ln\left(S_{0, i}/K_i\right)+\frac{1}{2}P_{11}\left(T_i\right)}{\sqrt{P_{11}\left(T_i\right)}} \quad
d_{2, i} = d_{1, i}-\sqrt{P_{11}\left(T_i\right)},
\end{align}
where $r=0$, $T_i$, $S_{0, i}$ and $K_i$, defined in \eqref{eq:bl-price}, are given by the option contract and $P_{11}$, defined in \eqref{eq:analytical-solution}, depends on the model parameters $\sigma_1, \sigma_2$ and $\lambda$. 
We want the error between the model-predicted prices and the real market prices to be minimal according to a least squares criterion: we thus define the loss function
\begin{equation}
L\left(\sigma_1, \sigma_2, \lambda\right) = \sum_{i=1}^{n}\left(O_{i}-\hat{O}_{i}\left(\sigma_1,\sigma_2,\lambda\right)\right)^{2}.
\end{equation}
The market calibration can then be translated into the optimization problem
\begin{subequations}
\begin{align}
\sigma_1^*, \sigma_2^*, \lambda^* &= \arg \min_{\sigma_1, \sigma_2, \lambda}L\left(\sigma_1, \sigma_2, \lambda\right)\\
\text{subject to:} & \; \lambda \geq 0, \; \sigma_1 \geq 0, \; \sigma_2 \geq 0,
\end{align}
\label{eq:opt-problem}
\end{subequations} 
\noindent
which is non-convex. This can be intuitively understood from the non convexity of the Black formula with respect to the implied volatility or, alternatively, be shown with a counter example - see \cref{fig:loss-convexity}. A general non-convex solver, such as L-BFGS-B, a variant of the quasi-Newton method of Broyden, Fletcher, Goldfarb, and Shanno (BFGS) which accounts for box constraints \cite{zhu1997algorithm, nocedal2006numerical}, can be adopted when either the numerical or the analytical solutions are used to conpute $P_{11}$. However, the analytical solution enables the adoption of different solvers, such as trust region algorithms, which require first- and second-order derivatives. The computation of the gradient and the Hessian matrix of $L$ is tedious and error-prone, but can be performed by a library for automatic differentiation.

As all the non-convex optimization routines call for the repeated evaluation of pricing formula it is worth investigating how the numerical and the analytical solutions discussed in \cref{sec:variance-derivation} compare in terms of outcome and efficiency.

\begin{figure}
\centering
\begin{tikzpicture}
\begin{axis}[
	xlabel=$\sigma_2$,
	ylabel=$\Tilde{L}(\sigma_2)$,
	height=5cm,
    grid=major,
    grid style={thin, dotted, black!20},
]
\addplot[blue, mark=none] table [x=sy,y=L,col sep=comma]{"data/loss_convexity.csv"};
\end{axis}
\end{tikzpicture}
\caption{Consider the function $\Tilde{L}(\sigma_2) = L(\lambda, \sigma_1, \sigma_1)$, with $\lambda=1$ and $\sigma_1=1$ computed on a single option. The plot shows that $\Tilde{L}$ is not convex on the segment $\sigma_2 \in [10^{-4}, 20]$. As there exist a line on which the restriction of $L$ is not convex, we conclude that $L$ is non-convex.}
\label{fig:loss-convexity}
\end{figure}

\subsection{Experimental setup}
We considered European options from TTF gas and EEX electricity markets due to their high liquidity. The dataset included options on monthly futures listed in 57 consecutive days, between \nth{1} November 2017 and \nth{25} January 2018. The number of different delivery periods was 25 for TTF and 7 for EEX. The number of options changed daily, ranging between 430 and 596 for EEX and between 342 and 720 for TTF. 

We wanted to simulate the real-world scenario, where models are calibrated on listed options and used to price other instruments. Thus, on each trading day, 70\% of the available options were put in the set $\mathcal{C}$ used for calibration, while the remaining 30\% were placed in the set $\mathcal{V}$ to validate the model accuracy. Even though a study on the effectiveness of LMR-GW in interpolating option prices is not in the scope of this paper, we still need to guarantee that the calibration procedure achieves good results. To this goal, we choose the Mean Absolute Error (MAE) on the validation set as a performance metric:
\begin{equation}
    \text{MAE} = \frac{1}{|\mathcal{V}|}\sum_{i \in \mathcal{V}} \left| O_{i}-\hat{O}_{i}\right|,
\end{equation}
where $|\mathcal{V}|$ denotes the cardinality of $\mathcal{V}$.

In order to prove the effectiveness of market calibration, historical calibration, performed with a discrete Kalman filter \cite{MartinBarlow2004, Schwartz2000} is considered for comparison. The optimization problem \eqref{eq:opt-problem} was solved using both the L-BFGS-B solver and the trust-region solver provided by SciPy. The trust-region solver could only be applied when the analytical solution of the variance was adopted and was fed with the gradient and the Hessian computed by the automatic differentiation library Jax \cite{jax2018github}

\begin{table}
\centering
\begin{tabular}{l r r}
\hline 
{} & \textbf{EEX} & \textbf{TFF} \\
\hline 
Days & 57 & 57 \\
Average number of options per day & 500 & 530 \\
CPU time - analytical solution with L-BFGS-B [s] & 55 & 66 \\
CPU time - analytical solution with trust region [s] & 240 & 279 \\
CPU time - numerical solution with L-BFGS-B [s] & 632 & 1058 \\
MAE on $\mathcal{V}$ - historical calibration [\euro/MWh] & 0.354 & 0.276 \\
MAE on $\mathcal{V}$ - market calibration [\euro/MWh] & 0.072 & 0.045 \\
\hline
\end{tabular}
\caption{Dataset, calibration accuracy, and CPU time.}
\label{tab:market-calibration}
\end{table}

\subsection{Experimental results} \label{sec:experimental-results}
\subsubsection*{Computational efficiency}
Complete results are given in \cref{tab:market-calibration}. On the hardware described in \cref{subsec:num-and-analytical-comparison}, the calibration of LMR-GW models with analytical variance took about 2 minutes in net CPU time, while slightly more than 28 minutes were required using the numerical solution. In our case study, the trust region solver with first and second-order derivatives could not achieve any improvement is computational speed. The speedup due to the analytical solution is about 14, significantly lower than the figures presented in \cref{subsec:num-and-analytical-comparison}. The discrepancy is explained by the fact that only a fraction of the time required by the optimization process is spent evaluating the variance. Still, a similar gain in performance is significant in a business context, as models on tens of markets are usually calibrated each day.

\subsubsection*{Calibration output}
The average relative absolute difference in predicted prices between the models featuring the analytical and the numerical solution was in the order of $10^{-10}$, thus suggesting that the two methods are indeed equivalent in terms of outcome. 

In order to show the difference between the historical and the market calibration, we consider the MAE reported in \cref{tab:market-calibration}. Moreover, we pick a sample day, \nth{28} November 2017, and we plot the prices predicted by the models for the options on TTF futures. As \cref{fig:calibration-gof} shows, the historical calibration implemented by the Kalman filter is not capable of capturing the market scenario and thus results inadequate for pricing tasks. This conclusion is in agreement with the discussion presented in \cref{sec:intro}.

\begin{figure}
	\centering
	\begin{subfigure}[b]{.48\linewidth}
		\includegraphics[width=.98\linewidth]{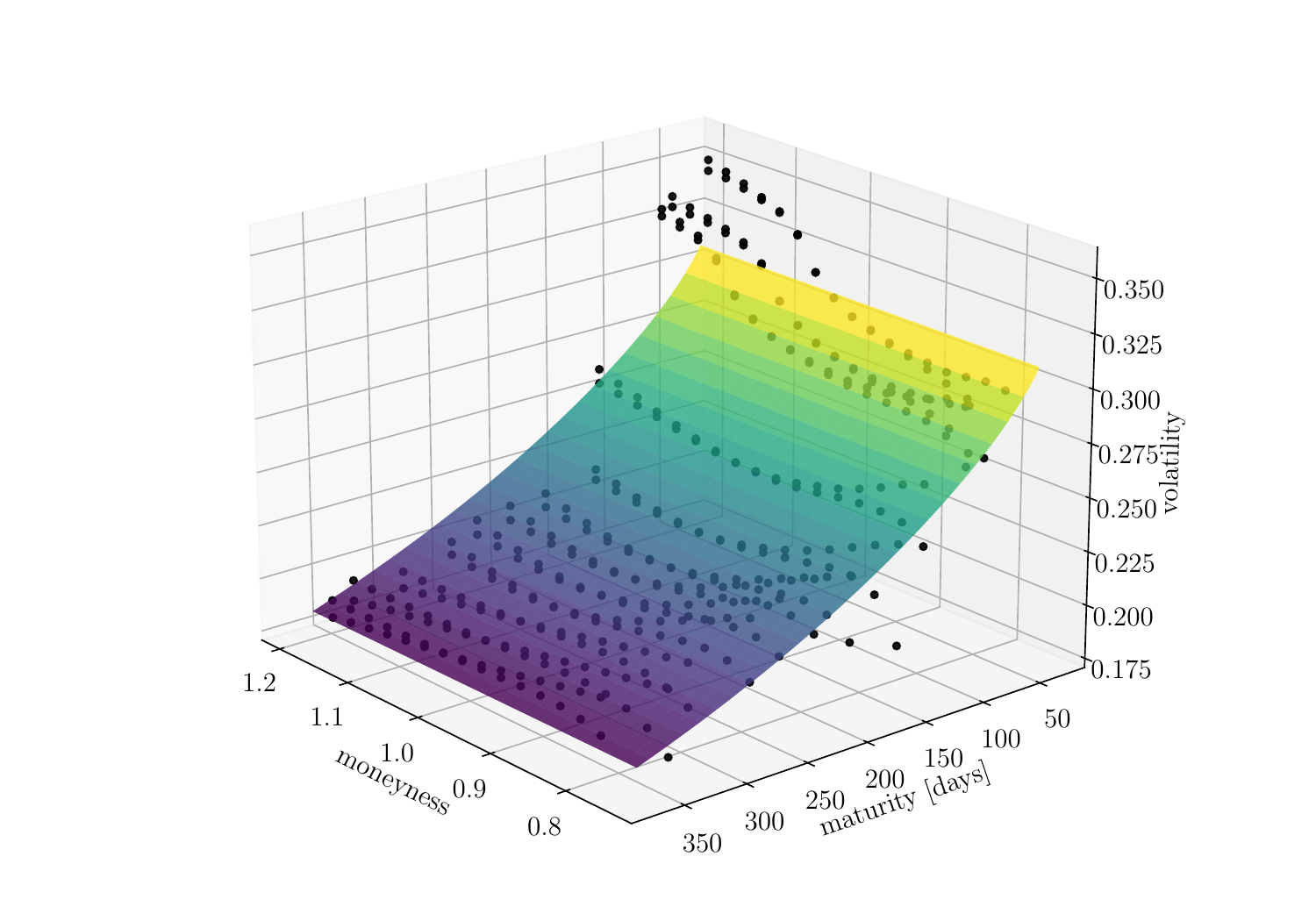}
		\caption{Ornstein-Uhlenbeck}
	\end{subfigure}
	\begin{subfigure}[b]{.48\linewidth}
		\includegraphics[width=.98\linewidth]{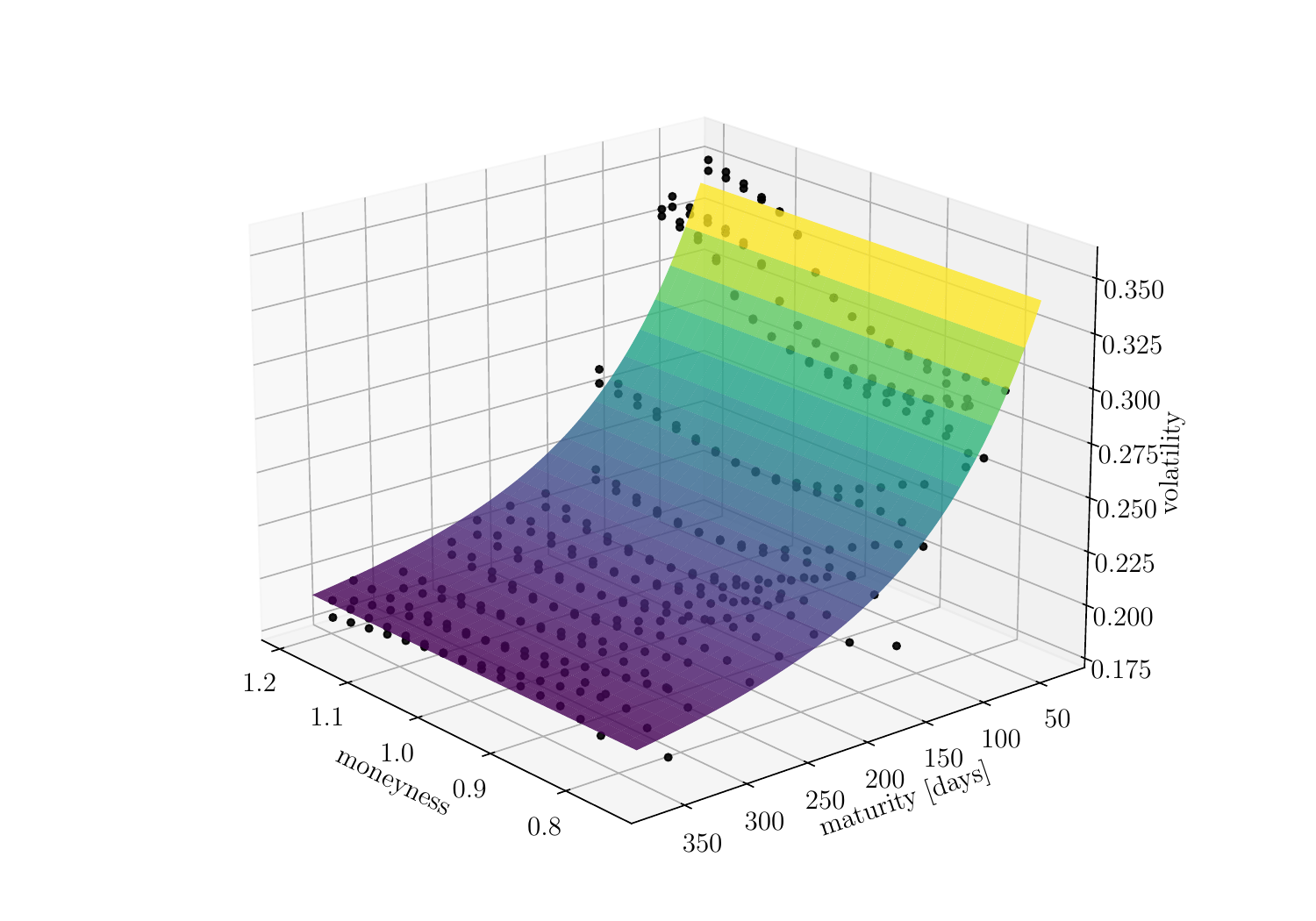}
		\caption{LMR-GW}
	\end{subfigure}
	\caption{Implied volatility against maturity and moneyness. The surface represents the volatility of the calibrated models, while the dots the implied volatility of listed options.\label{fig:Implied-volatility}}
\end{figure}

An example of the output of market calibration is depicted in \cref{fig:Implied-volatility}, where the implied volatility extracted from the fitted model is compared with the one of listed options. The implied volatility is the $\sigma$ parameter of the Black formula based on GBM - see \eqref{eq:bl-price} - and is an equivalent representation of the price of options. The chart displays options on monthly TTF futures as of \nth{28} November 2017 and a market-calibrated Ornstein-Uhlenbeck process is also included for reference and comparison. It is apparent that the calibration allows the model to fit well the implied volatility of listed options, with LMR-GW being more effective in capturing the Samuelson effect \cite{jaeck2016volatility} for very short maturities, as it can be seen from the slightly more pronounced curvature of the surface as a function of maturity. It shall be noted, though, that both models fail to capture "smile" effects, due to the lack of stochastic volatility. Such considerations match with the expectations based the literature and further certify the validity of the calibration procedure.

\begin{figure}
	\centering
	\begin{tikzpicture}
		\begin{axis}[
			height=5cm,
			grid=major,
			grid style={thin, dotted, black!20},
			xlabel={Predicted prices $\hat{O}_i$},
			ylabel={Actual prices $O_i$},
			legend pos=north west
			]
			\addplot[only marks,mark size=0.8pt,blue] table[x=price, y=lmrgw_price, col sep=comma] {"data/gof.csv"};
			\addlegendentry{Market calibration}
			\addplot[only marks,mark size=0.8pt,orange] table[x=price, y=hist_price, col sep=comma] {"data/gof.csv"};
			\addlegendentry{Historical calibration}
			\addplot[black] table[x=price, y=price, col sep=comma] {"data/gof.csv"};
		\end{axis}
	\end{tikzpicture}
	\caption{Goodness of fit for market and historical calibration for options in the validation set $\mathcal{V}$. A perfect model would result in all points lying on the bisector of the quadrant - the solid black line.}
	\label{fig:calibration-gof}
\end{figure}

\section{Conclusion}\label{sec:conclusion}

In this paper, we suggested an approach based on the Lyapunov equation that enables an easy derivation of the variance of a wide class of stochastic models. By plugging the variance into the Black formula, one can obtain pricing models more adherent to the behaviour of energy commodities. Moreover, we presented a detailed comparison between the computational performance of the analytical and the numerical solution of the Lyapunov equation, which may help practitioners to decide when it is worth investing effort in the derivation of explicit expressions. Finally, we showcased the practical implications of our study: the ready availability of analytical and numerical results for the variance of two-factor models finds applications in the market calibrations of stochastic processes. 

A Python implementation of the procedures described in the paper is available in an open-source package \cite{fabbiani2019vanilla, fabbiani2020vanilla}.

Future work may further expand the applications of the approach here presented, considering for example its usage in pricing spread options or deriving the Value at Risk (VaR) in risk models.

\appendix
\section{Application to Schwartz model} \label{appendix-schwartz}
We apply the Lyapunov equation to propose a simpler derivation for the variance of the linear two-factor model by Schwartz and Smith \cite{Schwartz2000}.
Let $s_t$ denote the price of the underlying at time $t$, $z_{\chi}$ and $z_{\xi}$ be standard Wiener processes such that $dz_{\chi}dz_{\xi} = \rho dt$. The original model is defined by:
\begin{subequations}
\begin{align}
	X(t) &= \ln\left(s(t)\right) \\
	X(t) &= \chi(t) + \xi(t) \\
	d\chi(t) &= -k\chi(t)dt + \sigma_{\chi}dz_{\chi}(t) \\
	d\xi(t) &= \mu_{\xi}dt + \sigma_{\xi}dz_{\xi}(t).
\end{align}
\end{subequations}{}
We write a state-space description of the model by defining:
\begin{equation}
A=\begin{bmatrix}
-k & 0 \\
0  & 0
\end{bmatrix} \quad
B=\begin{bmatrix}
\sigma_{\chi} & 0 \\
0 & \sigma_{\xi}
\end{bmatrix} \quad
C=\begin{bmatrix}
1 & 1
\end{bmatrix} \quad
S=\begin{bmatrix}
1 & \rho \\
\rho & 1 
\end{bmatrix}.
\end{equation}
It is now possible to run the computation presented in \cref{sec:variance-derivation} to derive the variance of the state. Let us define:
\begin{equation}
	x(t) = \begin{bmatrix}
	\chi(t) \\ 
	\xi(t)
	\end{bmatrix}, \quad
	P(t) = \mathrm{Var}[x(t)], \quad
	P_0 = P(0) = \mathbf{0}_{2 \times 2},
\end{equation}
where $\mathbf{0}_{2 \times 2}$ denotes a 2-by-2 null matrix. By applying \eqref{eq:analytical-solution}, one can compute:
\begin{align}
P\left(t\right)&=e^{At}P_{0}e^{A^{T}t}+\int_{0}^{t}e^{A(t-z)}BSB^{T}e^{A^{T}\left(t-z\right)}dz \nonumber \\
&=\begin{bmatrix}
\frac{\sigma_{\chi}}{2k}\left(1 - e^{-2kt}\right) & \frac{\rho \sigma_{\chi} \sigma_{\xi}}{k}\left(1 - e^{-kt}\right) \\
\frac{\rho \sigma_{\chi} \sigma_{\xi}}{k}\left(1 - e^{-kt}\right) & \sigma_{\xi}^2t 
\end{bmatrix}.
\end{align}
which matches Eq. 5b in Schwartz's work. Moreover, by applying \eqref{eq:output-variance}, one gets:
\begin{equation}
\mathrm{Var}\left[X(t)\right] = CP(t)C^T = \frac{\sigma_{\chi}}{2k}\left(1 - e^{-2kt}\right) +
2 \frac{\rho \sigma_{\chi} \sigma_{\xi}}{k}\left(1 - e^{-kt}\right) + \sigma_{\xi}^2t, 
\end{equation}
which is equation 6b in Schwartz's paper.

\bibliographystyle{unsrt}
\bibliography{references}

\end{document}